\documentclass[runningheads]{llncs}
\usepackage[T1]{fontenc}
\usepackage{graphicx}
\usepackage{url}
\usepackage{hyperref}
\usepackage{color}

\begin{document}
\title{Finite-Time Lyapunov Exponent Calculation on FPGA using High-Level Synthesis Tools}
\titlerunning{FTLE Calculation on FPGA using High-Level Synthesis Tools}

\author{Manuel de Castro\inst{1}\orcidID{0000-0003-3080-5136} \and \\
Roberto R. Osorio\inst{2}\orcidID{0000-0001-8768-2240} \and \\
Francisco J. Andújar\inst{1}\orcidID{0000-0001-8884-7334} \and \\
Rocío Carratalá-Sáez\inst{3}\orcidID{0000-0001-8409-2421} \and \\
Yuri Torres\inst{1}\orcidID{0000-0002-3037-3567} \and \\
Diego R. Llanos\inst{1}\orcidID{0000-0001-6240-9109} 
}

\authorrunning{De Castro, Osorio, Andújar, Carratalá-Sáez, Torres and Llanos}
\institute{Department of Computer Science, Universidad de Valladolid, Spain \\
\email{\{manuel,fandujarm,yuri.torres,diego\}@infor.uva.es} \and
CITIC, Computer Architecture Group, Universidade da Coruña, Spain\\
\email{roberto.osorio@udc.es} \and
Dpto. Ingeniería y Ciencia de los Computadores, Universitat Jaume I, Spain\\
\email{rcarrata@uji.es}
}

\maketitle              % typeset the header of the contribution
\begin{abstract}
As Field Programmable Gate Arrays (FPGAs) computing capabilities continue to grow, also does the interest on building scientific accelerators around them. Tools like Xilinx's High-Level Synthesis (HLS) help to bridge the gap between traditional high-level languages such as C and C++, and low-level hardware description languages such as VHDL and Verilog. In this report, we study the implementation of a fluid dynamics application, the Finite-Time Lyapunov Exponent (FTLE) calculation, on FPGA using HLS. We provide speed and resource-consumption results for 2- and 3-dimensional cases. 

\end{abstract}

\section{Introduction to HLS tools}
\label{sec:introduction}

For AMD/Xilinx FPGAs, a high-level synthesis language based on C++ is offered, called 
Vitis HLS (as part of the Vivado design tools). This high-level language allows the designer to focus on the most important 
aspects of hardware design, such as structure, parallelism, reuse, and pipelining. 
Other aspects such as arithmetic implementation, or retiming, are managed automatically 
by HLS.
In this way, the best implementation for both fixed point and floating point arithmetic 
functions is selected by the framework, even for transcendental functions, to meet 
the desired cycle length. With retiming, changing the target clock frequency is possible, 
and hardware will be pipelined accordingly. HLS will match the delays of different 
paths, inserting flip-flops for short delays, or RAM-based FIFOs if long delays are 
needed. All these tasks are complex and error-prone when carried out manually in 
VHDL or Verilog.

One key feature of Vitis HLS is that it retains a certain low-level control of 
what is happening internally to the FPGA. Thus, not only it is possible for the designer 
to guide the synthesis process, but the compilation reports and analysis tools included in Vitis
HLS can provide certain useful feed-back information. 

In the typical Vitis HLS workflow, designers are expected to distribute the code 
in functions and loops, specifying by means of \textit{pragmas} whether they must be pipelined 
or not. Additional \textit{pragmas} on input/output parameters allow the specification of 
how data are transmitted, while HLS will implement dataflow, FIFO, or random-access 
interfaces, inserting the right code and IP. In this way, and as an example, the 
function  \texttt{float product(float a, float b)\{return a*b;\}} would implement 
a floating point product every \textit{L} cycles, \textit{L} being the latency of 
the multiplier. However, adding \texttt{\#pragma HLS pipeline II = 1}, would fully 
pipeline the multiplier and offer 1 result per cycle; while \texttt{\#pragma HLS 
INTERFACE m\_axi port = a} would specify that port \texttt{a} is connected to RAM 
memory, and HLS would calculate the latency of all accesses based on the characteristics 
of a particular device and board. While SYCL, and specially OpenCL, provide features 
similar to these described for programming FPGAs, they are fewer in quantity and 
higher in level, working on more abstract constructs.

In summary, HLS allows much of the flexibility of encoding with VHDL or Verilog, 
without the burden of instantiating, connecting, and pipelining components. This 
reduces design time and coding errors. Furthermore, many device specific components, 
such as memory interfaces, are dealt with automatically, improving code portability. 
Vitis HLS is the right tool for optimizing a specific algorithm for Xilinx FPGAs, 
losing little expressiveness compared to VHDL or Verilog.

If higher level programming is required, targeting a broader choice of platforms, 
then Xilinx also offers the Vitis Unified Software Platform (USP) software environment. Vitis USP enables the development 
of embedded software and accelerated applications on heterogeneous Xilinx platforms, 
including FPGAs, SoCs, and Versal ACAPs. It provides a unified programming model for 
accelerating Edge, Cloud, and Hybrid computing applications. In addition, OpenCL is 
supported in Vitis USP, providing code portability. The price to pay is less 
control on the implementation when compared to Vitis HLS and, even more so, to VHDL 
and Verilog.

\subsection{Performance degradation vs portability in HLS frameworks}

HLS tools are known to introduce some overheads in resource consumption, specially 
those of higher abstraction, but they bring significant advantages when compared 
to VHDL or Verilog. In the case of lower-level frameworks particularly, such as Vitis 
HLS, it is important to note that little to none time savings are possible by writing 
the equivalent HDL code by hand, as an engineer would basically create the circuit 
instantiating the same IP blocks used by Vitis HLS. 

Although some improvements can be achieved using custom implementations, engineers 
must assess if those gains in area or speed are significant in the context of each 
application. In this line, the best known effort in the community is probably 
{FloPoCo}~\cite{FloPoCo_site,FloPoCo}.

It is also necessary to consider that additional performance improvements can be 
obtained by HLS tools if a slight precision degradation is allowed. These improvements 
would be larger than those obtained by redesigning the whole algorithm in an HDL 
language. As an example, Goldschmidt division and square root algorithms~\cite{GoldschmidtBruguera} 
might have shorter latency than the SRT algorithms~\cite{srt_algorithms}, which are 
widely used in many microprocessor implementations; but differences may appear in 
the bit in the last position. Fusing operations may also allow for some area and 
latency reductions, with Vitis HLS being able to efficiently exploit fused multiply-add 
(FMA) operations. Finally, it is possible to write custom arithmetic operations in 
C++ and Vitis HLS will produce efficient HDL code, as shown in~\cite{cube}, where 
the cube function is optimized using Vitis HLS to use less resources and reduce 
latency.

\section{Developing an implementation with Vitis HLS}
\label{sec:naive_theory}

Vitis HLS allows FPGA solutions to be developed with an approach close to the low-level 
details of FPGAs. This approach allows the developer 
to evaluate designs from information gathered at compile time and presented in the 
compilation reports, with a higher level of depth and detail than other frameworks such as SYCL and OpenCL. 
Therefore, Vitis HLS was used to develop a naïve implementation of the FTLE (finite-time Lyapunov exponent) 2D and 
3D kernels from the UVaFTLE application~\cite{rocio_uvaftle}, used in Fluid Dynamics, to identify issues which hinder performance, such as bottlenecks. 

\subsection{Baseline approach}

First, a naïve porting of the original FTLE kernels was considered. However, Vitis 
HLS failed to provide performance estimates (working clock frequency, initiation 
interval, and latency) for this naïve code without any transformation. This suggests 
that the unmodified kernel is too complex to efficiently synthesize into FPGA devices. 
After some work, targeting a 4-bank memory architecture (which is common in modern boards), 
HLS is able to schedule one 
memory read operation per cycle. In the case of the 2D algorithm, this would mean 
a maximum throughput of 0.25 points per clock cycle, or 125 million points per second. 
For the 3D algorithm, 0.166 points per cycle, or 59.5 million points per second. 
This implementation clearly underutilizes the bandwidth of double-data-rate memory. 
A different approach is needed to synthesize an efficient FTLE hardware architecture.

Therefore, an implementation comprising only the core of the FTLE computations was 
tested. C code was written assuming that all input data are available every cycle, 
and \textit{pragma} directives were set for fully pipelined architectures. The targeted FPGA 
was a Virtex Ultrascale+ (model xcvu11p-fsgd2104-3-e), a high-end FPGA. 
The resulting circuits were analyzed using the schedule viewer included in Vitis 
HLS to confirm that throughput had been maximized, and minimal latency was achieved. 
Maximum frequency, latency, required input bandwidth, and resource consumption are shown in Table~\ref{tab:roberto_results_hls}. Required input bandwidth is expressed for both the maximum operating frequency reported by the compiler and for 300 MHz. The reason for including results for 300 MHz is to provide a comparison point for a modest frequency that many other devices can reach, not only high-end ones. Resource usage actually quite modest: just 2\% of the available resources in the targeted device. Only in the 3D case the amount of DSP blocks is higher, taking 10\% of the total. 

% TODO: include resource usage?

\begin{table}[]
\centering
\begin{tabular}{c|c|c}
%\cmidrule{2-3}
 & \textbf{2D} & \textbf{3D} \\ \hline
{\textbf{Max Freq (MHz)}} & 500 & 357 \\ \hline
{\textbf{Latency / cycles}} & 264 & 421 \\ \hline
{\textbf{Input bandwidth (bits/cycle)}} & 768+128 & 1152+192 \\ \hline
{\textbf{Input bandwidth for max freq (GB/s)}} & 48+8 & 51.4+8.6 \\ \hline
{\textbf{Input bandwidth for 300 MHz (GB/s)}} & 28.8+4.8 & 43.2+7.2 \\ \hline
{\textbf{LUT}} & 29323 & 134519 \\ \hline
{\textbf{LUTRAM}} & 1797 & 5679 \\ \hline
{\textbf{FF}} & 49677 & 139912 \\ \hline
{\textbf{DSP}} & 250 & 1012 \\ \hline
{\textbf{BRAM}} & 0 & 1 \\ \hline
{\textbf{Power consumption}} & 8.1 W & 21.17 W \\ \hline
\end{tabular}
\vspace{\baselineskip}
\caption{Reported synthesis data for the Vitis HLS implementation of the FTLE 
kernels. \textit{Input bandwidth} shows the desired figures expressed as \textit{floating-point 
data bandwidth + neighbor indexes bandwidth}.}
\label{tab:roberto_results_hls}
\end{table}

As it can be seen, such large bandwidth requirements suggest that the throughput will 
be limited in most platforms by the real available bandwidth. Without bandwidth limitations, the fully-pipelined 2D architecture achieves 24.6 GFLOPS, and the 3D one 61.8 GFLOPS. We will assume that 
the circuit receives four indexes at each iteration, or six for the 3D case, pointing 
to the neighboring points of the current one. Those values are always read in the 
same order, so either a FIFO or an actual RAM could be used to store them. Using 
32-bit integers as indexes, 128 or 192 bits are read every iteration. Next, it is 
necessary to read from memory some coordinate and flowmap values for those neighboring 
points. In the 2D case, 12 values must be read. In the 3D case, 18 values must be 
read. Considering double-precision arithmetic, 728 and 1152 bits should be read from 
random addresses per cycle.

The implemented pipeline for the FTLE core has an initiation interval of 1, which 
implies that, if the memory bandwidth requirements are satisfied, the application 
would be able to process one point per clock cycle. For example, with a frequency 
of 500 MHz, achievable by the targeted FPGA board for the 2D design, the throughput 
would be 500 million points per second. With a frequency of 357 MHz, achievable by 
the targeted FPGA board for the 3D design, the throughput would be 357 million points 
per second. What follows is an analysis of how close it is possible to get to this 
maximum performance when synthesizing the whole FTLE computation to the FPGA, which 
must include the logic to read these data from global memory.

The first approximation is considering that all the data are in DDR memory. The peak 
bandwidth for one module of DDR4-2400 is 19.2 GB/s, and 21.3 GB/s for DDR4-2666. 
Adding more modules in parallel multiplies the bandwidth accordingly. Historically, 
Xilinx has advised against going beyond dual memory because of the increase in power 
consumption; although they now have some device models with four modules, as the 
DDR4 memory power consumption is lower than that of DDR2 and DDR3. Thus, the 
particular case of four DDR4-2400 modules is also considered. Therefore, we could expect up to 38.4 GB/s or 
42.6 GB/s for 2 channels of DDR4-2400 and DDR4-2666, respectively; and 76.8 GB/s 
for 4 channels of DDR4-2400, but only if a predictable access pattern is used.
For random access, the bandwidth will be lower.

In a second approximation, the data are first loaded into HBM (high-bandwidth memory)~\cite{HBM_BSC}. 
HBM provides a high bandwidth and low power consumption, as it is implemented inside 
the FPGA packaging, although in a different die. Xilinx advertises 230 GB/s for one 
stack, and 460 GB/s using two stacks. These figures are possible if concurrent accesses 
do not incur in bus conflicts; however, as two stacks provide almost 8 times the 
highest desired bandwidth for our target application, this should not be an issue.

% TODO: specify explicit Vivado HLS used. Explain Vivado HLS vs Vitis HLS
HBM is being used in many high-performance computing applications~\cite{Serpens}, 
and helps to overcome memory bandwidth hurdles that limit the implementation of many 
applications on FPGAs. Unfortunately, we were not able to infer HBM memory within HLS using pragmas. 
%as Xilinx seems to focus on Vitis as the preferred working environment for its newer devices and boards.
At the current point, more knowledge is needed to instantiate HBM at high level or, as a last resource, 
to manually instantiate and configure it using VHDL. Table~\ref{tab:bandwidth_performance} 
compares the peak bandwidth of the application if executed on a system with the discussed 
different memory technologies.

\begin{table}[]
\centering
\resizebox{\textwidth}{!}{%
\begin{tabular}{c|c|c|c|c|c}
%\cline{2-6}
 & \textbf{Absolute BW} & \textbf{2D max freq} & \textbf{3D max freq} & \textbf{2D 
300 MHz} & \textbf{3D 300 MHz} \\ \hline
\multicolumn{1}{c|}{\textbf{Desired bandwidth}} & & 56 (100\%) & 60 (100\%)  & 33.6 
(100\%) & 50.4 (100\%) \\ \hline
\multicolumn{1}{c|}{\textbf{1 channel DDR4-2400}} & 19.2 & 34\% & 32\% & 57\% & 38\% 
\\ \hline
\multicolumn{1}{c|}{\textbf{1 channel DDR4-2666}} & 21.3 & 38\% & 36\% & 63\% & 59\% 
\\ \hline
\multicolumn{1}{c|}{\textbf{2 channel DDR4-2400}} & 38.4 & 69\% & 64\% & 114\% & 
76\% \\ \hline
\multicolumn{1}{c|}{\textbf{2 channel DDR4-2666}} & 42.6 & 76\% & 71\% & 127\% & 
85\% \\ \hline
\multicolumn{1}{c|}{\textbf{4 channel DDR4-2400}} & 76.8 & 137\% & 128\% & 229\% 
& 152\% \\ \hline
\multicolumn{1}{c|}{\textbf{1 stack HBM}} & 230 & 410\% & 383\% & 685\% & 456\% \\ 
\hline
\multicolumn{1}{c|}{\textbf{2 stack HBM}} & 460 & 820\% & 767\% & 1369\% & 912\% 
\\ \hline
\end{tabular}%
}
\vspace{\baselineskip}
\caption{Peak achievable bandwidth using different memory technologies, both in absolute 
terms and relative to the desired bandwidth. Absolute bandwidth is expressed in GB/s.}
\label{tab:bandwidth_performance}
\end{table}

The rows of table~\ref{tab:bandwidth_performance} presenting data for the DDR4-2400 
memory provide insight into the theoretical bandwidths achievable by many common data-center FPGAs
%used in the experimental 
%evaluations, Intel Stratix 10 FPGAs (PAC D5005)
, under different scenarios. In an 
ideal scenario, all four banks would be efficiently used, and the achieved bandwidth 
would be enough to efficiently compute the FTLE without stalls. Nevertheless, that 
scenario could only be achieved with optimal access patterns. Any introduction of 
irregularity or unbalance in the access to global memory can significantly reduce 
the effective bandwidth, as explored in~\cite{zohouri_memory}. If two banks are efficiently 
used, an FPGA would be able to compute the FTLE of 2D inputs without stalls 
at 300 MHz, but not for 3D inputs.

\subsection{Decoupling the problem}
\label{sec:optimized_theory}

In the previous section, we identified the original FTLE algorithm as a memory-bound 
code. The core computation of the algorithm can be efficiently implemented as a hardware 
pipeline, which any modern FPGA should be able to implement with an initiation interval 
of 1 (i.e. able to produce one FTLE result per clock cycle). Nevertheless, to perform 
this core computation, the neighbors of each of the points must be located beforehand, 
which constitutes the memory-intensive section of the algorithm. The neighbor determination 
presents a highly irregular memory access pattern, with very low data reutilization, 
which makes leveraging FPGA caching techniques unfeasible in any meaningful way. 
Thus, all data accesses must be issued to global memory. Given the relatively low global memory 
bandwidth of DDR-based FPGAs, which is specially low when 
the program is only able to leverage one DDR4 bank, this neighbor determination section 
constitutes the bottleneck of the algorithm on FPGAs. To make efficient FPGA FTLE 
architectures, this pressure on the global memory bandwidth should be alleviated.

We propose an FPGA-optimized algorithm for the FTLE in which the neighbors for each 
point are pre-computed and stored in a regular list of point indexes, i.e. integers.
This list is then fed into the core FTLE computation. For each point of the computation, 
there are exactly 4 or 6 indexes in the neighbors list, corresponding to the neighbors 
in 2D or 3D space, respectively. For irregular meshes, where not all points have 
the maximum number of neighbors, the absence of one of the neighbors is encoded with 
the value -1, thus being easily handled in the code while still preserving the regularity 
of the list and its corresponding access pattern. By precomputing the neighbors list, 
we allow the new FTLE kernel both to reduce memory accesses, and to present a regular 
access pattern.

% TODO: remove this paragraph? Or make significant
Certain algorithmic simplifications are also performed to the gradient computations, 
to help the compiler produce more efficient hardware. Some of these are minor optimizations 
that could be automatically detected by the compiler (e.g., removal of branches never 
taken, or merging of similar branch cases), but we want to make sure that they are 
performed.

We propose the neighbors list should be precomputed on the CPU, instead of on the 
FPGA. The main reason being that we know the determination of neighbors achieves 
a poor performance on FPGAs, due to the memory issues discussed earlier. From a hardware-software 
codesign point of view, this approach is not only reasonable, but even desired, as 
the software and hardware components of the system perform the tasks that are optimal 
for them.

Furthermore, it is not unreasonable to think of a scenario in which the neighbors 
list can be provided as input to the application together with the points' data. This 
would completely remove the need for any extra computation on the CPU side of the 
application. Nevertheless, that scenario is not considered in this work, and we consider 
only cases where the inputs to the application are the ones used by the original 
implementation of UVaFTLE.

This new optimized algorithm is considerably simpler than the naïve one, so other synthesis tools
(e.g. OpenCL and SYCL)
should not encounter problems when attempting to synthesize a pipeline 
for it either. Additionally, the much simpler memory access pattern should alleviate the 
memory constraints of the naïve kernels, achieving a higher effective bandwidth. 
The performance of these new kernels will still be limited by the memory bandwidth 
of the device, with an inefficient memory data management resulting in pipeline stalls. 
In this version of the kernels, the memory data management is still implicitly relegated 
to the compiler, which can perform automatic optimizations by allocating different 
buffers to different memory banks.

\section{Conclusions}
\label{sec:conclusions}

In this work, we have presented implementation results for the UVaFTLE application on AMD/Xilinx FPGAs, and we have achieved a number of conclusions. First, non-floating point calculations are better off-loaded to the host processor, as those are highly irregular, and a microprocessor is better suited to carry them out thanks to its higher clock frequency. Second, the implementation performance is limited by memory bandwidth, and maximum performance is only possible by using architectures such as HBM. Third, the core of the application, consisting of floating point calculations, can be efficiently implemented as a deep pipeline with a performance of several GFLOPS: Approximately 24.6 GFLOPS for 2D computations, and 61.8 GFLOPS for 3D computations.

% Trabajo futuro.
Future work includes performing experimental evaluations of the Vitis HLS codes 
on Xilinx FPGAs, studying the optimal approach to accelerate the determination of
the list of neighbors for each point, and developing solutions which leverage HBM
memory.

\section{Acknowledgments}

This work was supported in part by: The Spanish Ministerio de Ciencia e Innovación 
and by the European Regional Development Fund (ERDF) program of the European Union, 
under Grant PID2022-142292NB-I00 (NATASHA Project); and in part by the Junta de Castilla 
y León - FEDER Grants, under Grant VA226P20 (PROPHET-2 Project), Junta de Castilla 
y León, Spain. This work was also supported in part by grant TED2021–130367B–I00, 
funded by MCIN/AEI/10.13039/ 501100011033 and by “European Union NextGenerationEU/PRTR”, 
and by grant PID2022-136435NB-I00, funded by MCIN/AEI/ 10.13039/501100011033 and 
by “ERDF A way of making Europe”, EU. Manuel de Castro has been supported by
Spanish Ministerio de Ciencia, Innovación y Universidades, through “Ayudas para la
Formación de Profesorado Universitario FPU 2022”.

\subsection*{Disclosure of Interests}
The authors have no competing interests that
might be perceived to influence the results and/or discussion reported in this paper.

\subsection*{Availability of data and materials}
The source codes and compilation reports generated during the development of this 
work are freely available on the following repository: \url{https://github.com/uva-trasgo/uvaftle/tree/fpga}

\bibliographystyle{splncs04}
\bibliography{bibliography}% common bib file

\begin{thebibliography}{1}
\providecommand{\url}[1]{\texttt{#1}}
\providecommand{\urlprefix}{URL }
\providecommand{\doi}[1]{https://doi.org/#1}

\bibitem{FloPoCo_site}
{FloPoCo} project website. \url{http://www.flopoco.org} (October 2022)

\bibitem{HBM_BSC}
Asifuzzaman, K., et~al.: Demystifying the characteristics of high bandwidth
  memory for real-time systems. In: 2021 IEEE/ACM International Conference On
  Computer Aided Design (ICCAD). pp.~1--9 (2021).
  \doi{10.1109/ICCAD51958.2021.9643473}

\bibitem{rocio_uvaftle}
Carratalá-Sáez, R., et~al.: {UVaFTLE: Lagrangian finite time Lyapunov
  exponent extraction for fluid dynamic application}. {The Journal of
  Supercomputing}  \textbf{79},  9635--9665 (2023).
  \doi{10.1007/s11227-022-05017-x}

\bibitem{FloPoCo}
de~Dinechin, F., Pasca, B.: Designing custom arithmetic data paths with
  {FloPoCo}. {IEEE} Design \& Test of Computers  \textbf{28}(4),  18--27 (Jul
  2011)

\bibitem{srt_algorithms}
Harris, D., Oberman, S., Horowitz, M.: Srt division architectures and
  implementations. In: Proceedings 13th IEEE Sympsoium on Computer Arithmetic.
  pp. 18--25 (1997). \doi{10.1109/ARITH.1997.614875}

\bibitem{cube}
Osorio, R.R.: Floating point calculation of the cube function on fpgas. IEEE
  Transactions on Parallel and Distributed Systems  \textbf{34}(1),  372--382
  (2023). \doi{10.1109/TPDS.2022.3220039}

\bibitem{GoldschmidtBruguera}
Pi{\~n}eiro, J.A., Bruguera, J.D.: High-speed double-precision computation of
  reciprocal, division, square root and inverse square root. IEEE Trans.
  Computers  \textbf{51},  1377--1388 (2002)

\bibitem{Serpens}
Song, L., Chi, Y., Guo, L., Cong, J.: Serpens: A high bandwidth memory based
  accelerator for general-purpose sparse matrix-vector multiplication. In:
  Proceedings of the 59th ACM/IEEE Design Automation Conference. p. 211–216.
  Association for Computing Machinery, New York, NY, USA (2022).
  \doi{10.1145/3489517.3530420}

\bibitem{zohouri_memory}
Zohouri, H.R., Matsuoka, S.: The memory controller wall: Benchmarking the intel
  fpga sdk for opencl memory interface. In: 2019 H2RC. pp. 11--18.
  \doi{10.1109/H2RC49586.2019.00007}

\end{thebibliography}

\end{document}